# Exploiting a Shortcoming of Coupled-Cluster Theory: The Extent of Non-Hermiticity as a Diagnostic Indicator of Computational Accuracy

Kaila E. Weflen, Megan R. Bentley, James H. Thorpe, Peter R. Franke, Jan M. L. Martin,* Devin A. Matthews, and John F. Stanton

**ABSTRACT:** The fundamental non-Hermitian nature of the forms of the coupled-cluster (CC) theory widely used in quantum chemistry has usually been viewed as a negative, but the present paper shows how this can be used to an advantage. Specifically, the non-symmetric nature of the reduced one-particle density matrix (in the molecular orbital basis) is advocated as a diagnostic indicator of computational quality. In the limit of the full coupled-cluster theory [which is equivalent to full configuration interaction (FCI)], the electronic wave function and correlation energy are exact within a given one-particle basis set, and the symmetric character of the exact density matrix is recovered. The extent of the density matrix asymmetry is shown to provide a measure of "how difficult the problem is" (like the well-known $T_1$ diagnostic), but its variation with the level of theory also gives information about "how well this particular method works", irrespective of the difficulty of the problem at hand. The proposed diagnostic is described and applied to a select group of small molecules, and an example of its overall utility for the practicing quantum chemist is illustrated through its application to the beryllium dimer ($Be_2$). Future application of this idea to excited states, open-shell systems, and symmetry-breaking problems and an extension of the method to the two-particle density are then proposed.

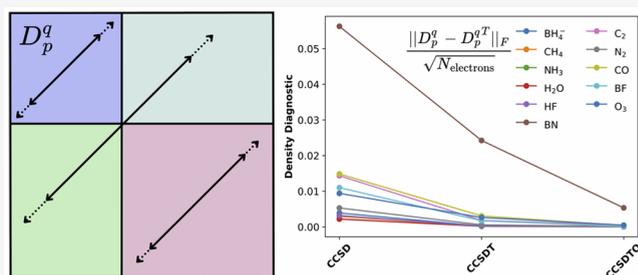

The tremendous success of quantum chemistry is such that few experimental chemists have escaped its influence. The emergence of practical and accurate methods within the computationally efficient density functional theory (DFT)[1,2] has taken the influence of the field from small- and medium-sized molecules to the realm of materials science and meaningful biological application. Behind the density functional theory in importance and impact is the coupled-cluster (CC) theory, which was imported from nuclear physics to quantum chemistry by Čižek nearly 60 years ago.[3] Unlike density functional approaches, the CC theory is systematically improvable and capable of providing results having sufficient accuracy to be meaningfully compared to (or used to accurately predict) experimental results in a wide regime of spectroscopic, thermochemical, and kinetic realms. The groundbreaking work of Neese and co-workers[4] has extended the CC theory to a range of problems that is beginning to be near that of the DFT, and developments in both areas, DFT and CC methods, are enduring and active areas of research in the theoretical chemistry community.

As it is systematically improvable, the CC theory can be exploited to produce higher- and higher-level approximate solutions to the electronic Schrödinger equation. However, this improvement comes at great computational cost. The traditional CCSD,[5] CCSDT,[6] and CCSDTQ[7] methods have computational scalings of $N^6$, $N^8$, and $N^{10}$, respectively ($N$ corresponds roughly to the size of the basis set used in the calculations); therefore, one ultimately must make a compromise between accuracy and cost. A question that necessarily arises in any quantum chemical study is "just how accurate are my results?". While comparisons to the experiment can offer insight, what does one do when the purpose of the calculation is predictive rather than the analysis of existing results?

Since the relatively early days of widespread applications of CC theory to molecular problems, the early to mid-1980s, pathologies have been noted. For example, CC theory is not variational, and many potential curves of diatomic molecules have been calculated from the near-equilibrium regime to large separation well along the highway to dissociation.[9] A comparison to full configuration interaction (CI) calculations done in small basis sets revealed potential energy curves diverging to energies below the exact values, along with other absurdities: for example, the prediction by the CCSD+T(CCSD) method[6] that ozone has a $C_s$ rather than $C_{2v}$ equilibrium geometry,[10] as well as, with CC2,[11] CC3,[12] and CC4[13] theories (all three of them), that the permanganate anion ($MnO_4^-$) spontaneously and preposterously dissociates into five atoms via a concerted fission of four chemical bonds.

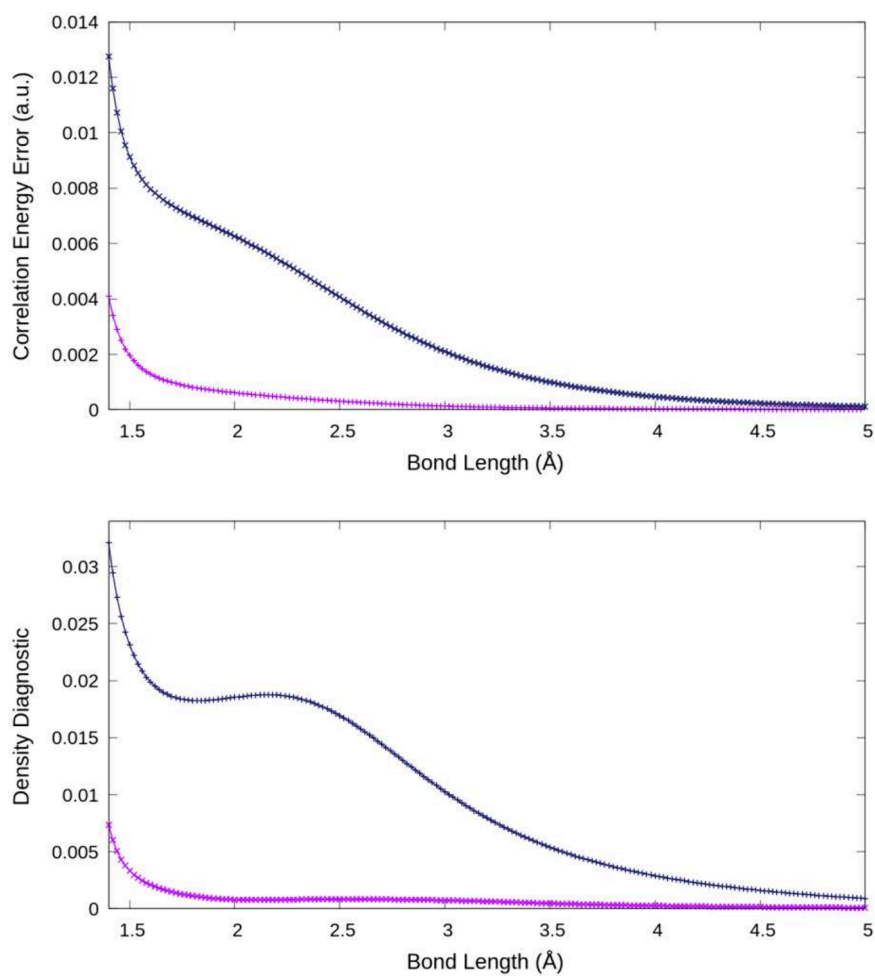

**Figure 1.** Density diagnostic (bottom) and correlation energy error (top) are shown as a function of $Be_2$ internuclear distances in the frozen-core approximation for the CCSD (dark blue) and CCSDT (magenta) levels of theory using the cc-pVDZ basis set.

In the above cases, the methods being discussed surely are not working very well. Seeing the need for simple insight into the accuracy prompted Lee and Taylor in 1989 to propose the $T_1$ diagnostic.[14] The said diagnostic indicator is extremely simple to obtain from a CC calculation and has been advocated as a measure of the "multireference character" (read computational difficulty) of molecular systems.[15] The widespread use of the $T_1$ diagnostic (customarily evaluated at the CCSD level) as a measure of difficulty testifies to its utility, despite some formal objections that can be made of this approach.[16] In the ensuing years, a vast number of other diagnostic indicators have been proposed (summarized and discussed at length in refs 17 and 18), all of which seem to provide some interesting insight into calculations done with CC methods.

This letter proposes another indicator of computational reliability. Although the need to evaluate the one-particle reduced density matrix (1PRDM) roughly doubles the computation time for an energy-only CC calculation (this step would be required anyhow for gradients), the new indicator has several attractive features. The most important is that the proposed diagnostic tells you not only how difficult a particular system is (often spoken as the extent of "multi-reference character") but also *how well a particular method does to solve the problem at hand*. It is this *second* property that makes the diagnostic unique and, in our view, more generally useful to quantum chemists than any existing measure of computational difficulty and quality.

As was first emphasized long ago by Arponen et al.,[19] electronic states given by the normal coupled-cluster approaches (CCD,[8] CCSD, CCSDT, ...) can be viewed as solutions to a non-Hermitian eigenvalue problem, in which the matrix $\bar{H} [\equiv \exp(-\hat{T})H \exp(\hat{T})]$ is diagonalized.[20] For the ground state, the right and left eigenvectors (the former of which is trivial) represent the unit vector and the so-called lambda state, well-known in the CC gradient theory. From this perspective, other states populating the Hilbert space, which again have left and (now non-trivial) right eigenvectors, represent other $n$-electron states of the system (this is the essence of the equation-of-motion CC method known as EOMEE-CC[21]), and straightforward extensions to different numbers of electrons give $n - 1$ electron states (EOMIP-CC), $n + 1$ electron states (EOMEA-CC), etc. In all such approaches,[21] the right and left electronic wave functions (the eigenvectors of $\bar{H}$ are distinct) form a biorthogonal set. For the ground state, which is the emphasis of this letter, the right and left electronic state wave functions are given by

$$|\Psi\rangle = \exp(\hat{T})|0\rangle \quad (1)$$

and

$$\langle\tilde{\Psi}| = \langle 0|(1 + \Lambda)\exp(-\hat{T}) \quad (2)$$

respectively. The equations above illustrate the non-Hermiticity of the CC theory; within a truncated CC method, the adjoint of the right-hand wave function is not proportional to the left-hand wave function, but this symmetry is restored in the exact (FCI) limit.[22]

A simple measure of this asymmetry is the one-particle reduced density matrix, elements of which are given by

$$D_p^q \equiv \langle 0|(1+\Lambda)\exp(-\hat{T})\{p^\dagger q\}\exp(\hat{T})|0\rangle \quad (3)$$

which can be inexpensively and easily computed in the course of any CC analytical gradient calculation. The extent of asymmetry of this quantity is the basis of the proposed diagnostic. Specifically, the following quantity:

$$\frac{\|D_p^q - D_p^{qT}\|_F}{\sqrt{N_{\text{electrons}}}} \quad (4)$$

where $\|\ \|_F$ represents the Frobenius norm of the antisymmetric contribution to the one-particle reduced density matrix, normalized by the square root of the total number of correlated electrons. Larger values of the diagnostic indicate that the wave function is farther from the full CI limit. Likewise, a reduction in the magnitude of the diagnostic accompanies improvement in the CC treatment, and it will ultimately vanish in the limit of the full configuration interaction. It should also be noted that the measure, as defined above, is size-intensive for any CC method, in the sense that the diagnostic calculated for $n$ identical and infinitely separated systems will be equal to that for the monomers. The numerical values of the quantity and its general behavior are illustrated in the following simple calculations.

Some important features of the proposed asymmetry diagnostic can be illustrated by the study of the beryllium dimer ($Be_2$), which is a molecule well-known in quantum chemistry.[23] Despite a vanishing formal bond order, this molecule is weakly bound and has been well-characterized by molecular spectroscopy. However, as the leading electronic configuration is $\sigma_{1s}^2\sigma_{1s}^{*2}\sigma_{2s}^2\sigma_{2s}^{*2}$, it is bound through electron correlation effects. For the simple calculations reported here, which use the modest cc-pVDZ basis[24] in the frozen-core approximation, the potential curve exhibits a minimum only when one goes beyond the CCSD level of theory. Figure 1 shows the quality of the potential for a range of internuclear distances, along with the associated values of the proposed diagnostic.[25] With regard to the latter, the following features are notable. First, for the highest level (CCSDTQ) calculation, the diagnostic vanishes for all distances, as this method provides an exact treatment (equivalent to FCI) for a system with four correlated electrons. However, it can be seen that both the CCSD and CCSDT diagnostics also vanish in the limit of a large internuclear distance. This arises from the size-extensive nature of CC theory, in which CCSD provides an exact treatment of this system at infinite separation.[26]

As one moves to smaller Be−Be distances, both the CCSD and CCSDT diagnostics gradually rise until shooting up quite rapidly at distances below 1.5 Å. In this region, there is a very strong configuration mixing between the $[\text{core}]\sigma_{2s}^2\sigma_{2s}^{*2}$ and $[\text{core}]\sigma_{2s}^2[\pi_{2px}^2 + \pi_{2py}^2]$ molecular orbital descriptions. Just slightly below the domain of the equilibrium geometry (2.60 and 2.56 Å with CCSDT and CCSDTQ, respectively), there is a weak (but largely insignificant) maximum seen in the CCSD diagnostic, after which it slowly decays as the atoms move out of the interaction region. The CCSDT diagnostic is considerably smaller, as is reflected in the binding energies: 137, 78, and 0 (purely repulsive) cm$^{-1}$ for CCSDTQ, CCSDT, and CCSD, respectively. Note that these are all well below the reasonably precise value of 839 cm$^{-1}$ recorded in ref 27 due to the effects of core correlation (neglected here for obvious instructive purposes) and basis set insufficiency in the present calculation.

All-electron calculations on the beryllium atom (see Table 1) again show that the diagnostic vanishes in the limit of an exact treatment of electron correlation (CCSDTQ) and that it is non-zero at the CCSD and CCSDT levels. At this point, without context, it should be noted that the magnitude of the diagnostic decreases monotonically and by more than 2 orders of magnitude as one makes the CCSD, CCSDT, and CCSDTQ progression. The maximum $T_2$ amplitude and the $T_1$ diagnostic are largely independent of the treatment of correlation, which is, of course, consistent with their "how hard is the problem" utility, while the proposed diagnostic also answers the "how well are we doing?" query.

Table 1. Harmonic Vibrational Frequencies (cm$^{-1}$), Equilibrium Bond Lengths (Å), and Density Asymmetry Diagnostic (DAD) Values for Beryllium-Based Molecules within the Molecular Test Suite at CCSD, CCSDT, and CCSDTQ Levels of Theory Using the cc-pVDZ Basis Set (All Electrons Correlated)[a]

| molecule | symmetry | property | CCSD | CCSDT | CCSDTQ |
|---|---|---|---|---|---|
| Be | | $T_1^{\text{diag}}$ | 0.01155 | | |
| | | $T_2^{\text{max}}$ | 0.14930 | | |
| | | DAD | 0.0002290 | 0.0000032 | 0 |
| BeO | $C_{\infty v}$ | $T_1^{\text{diag}}$ | 0.04326 | 0.04472 | 0.04493 |
| | | $T_2^{\text{max}}$ | 0.05942 | 0.06076 | 0.06095 |
| | | DAD | 0.0471888 | 0.0188874 | 0.0022748 |
| | | $\omega_1$ | 1493 | 1386 | 1363 |
| | | $r_{\text{Be-O}}$ | 1.34683 | 1.36697 | 1.36986 |
| BeOBe | $D_{\infty h}$ | $T_1^{\text{diag}}$ | 0.03414 | 0.03418 | 0.03419 |
| | | $T_2^{\text{max}}$ | 0.77547 | 0.77520 | 0.77515 |
| | | DAD | 0.0177736 | 0.0041337 | 0.0005552 |
| | | $\omega_1$ | 1013 | 994 | 991 |
| | | $\omega_2$ | 54 | 30 | 36 |
| | | $\omega_3$ | 1406 | 1376 | 1372 |
| | | $r_{\text{Be-O}}$ | 1.42833 | 1.43313 | 1.43400 |

[a]Associated $T_1$ diagnostics and maximum $T_2$ amplitudes at the CCSD level are provided for comparison.

A classic test case for multireference methods is the insertion of the Be atom into $H_2$.[28] We evaluated the insertion profile at the CCSD, CCSDT, and CCSDTQ (equivalent to full CI) levels with the cc-pVTZ basis set,[24] at the geometries of points A−J given in ref 28. As seen in Figure S1 and Table S1 of the Supporting Information, the difference between CCSD and CCSDTQ closely parallels the DAD(CCSD) diagnostic along

Table 2. Molecular Test Suite Diagnostic Values Using All-Electron (ae)-CCSD(T)/cc-pVTZ Reference Geometries[a]

| molecule | CCSD | CCSDT | CCSDTQ | $T_1^{\text{diag}}$ | $T_2^{\text{max}}$ |
|---|---|---|---|---|---|
| $BH_4^-$ | 0.0038991 | 0.0002657 | 0.0000184 | 0.00732 | 0.03913 |
| $CH_4$ | 0.0029586 | 0.0001753 | 0.0000197 | 0.00482 | 0.03249 |
| $NH_3$ | 0.0022457 | 0.0001472 | 0.0000226 | 0.00485 | 0.04761 |
| $H_2O$ | 0.0021959 | 0.0002036 | 0.0000195 | 0.00526 | 0.05138 |
| HF | 0.0032242 | 0.0001948 | 0.0000369 | 0.00499 | 0.04693 |
| BN | 0.0562656 | 0.0242503 | 0.0053528 | 0.06831 | 0.24557 |
| $C_2$ | 0.0143525 | 0.0018348 | 0.0004776 | 0.03099 | 0.31098 |
| $N_2$ | 0.0053171 | 0.0005790 | 0.0001290 | 0.00997 | 0.10451 |
| CO | 0.0148615 | 0.0030954 | 0.0004130 | 0.01653 | 0.07725 |
| BF | 0.0110149 | 0.0017292 | 0.0001465 | 0.01458 | 0.11171 |
| $O_3$ | 0.0094080 | 0.0026562 | 0.0004983 | 0.02384 | 0.21322 |

[a]The density diagnostic values are calculated at the theory level indicated in the table with a cc-pVDZ basis and all electrons correlated. The $T_1$ diagnostic and maximum $T_2$ amplitudes at the CCSD level for each species are included for comparison.

the reaction profile, and the same holds for CCSDT vs DAD(CCSDT).

Other less trivial examples are provided by the oxides BeO and BeOBe. The former is well-known to present a difficult example in quantum chemistry,[29] and the latter, recently studied and characterized experimentally by the Heaven group at Emory University,[30] carries with it an extensive "multireference" character, as is apparent from its largest $T_2$ excitation amplitude (see Table 1). While the $T_1$ diagnostics for both molecules are similar, the enormous highest occupied molecular orbital (HOMO) → lowest unoccupied molecular orbital (LUMO) double-excitation amplitude of BeOBe should give pause to any practitioner of quantum chemistry. When one sees an amplitude of this size, the single Slater determinant reference upon which "normal" CC methods are based is called into question. In contrast, BeO appears to be largely single-reference, and one might expect that this diatomic is treated better than BeOBe. However, the proposed diagnostic tells a different story. While both are much larger than the values for the single beryllium atom (as one expects), the values for the highly multireference BeOBe example is actually smaller than that for BeO. The equilibrium structural parameters and harmonic vibrational frequencies for these species, also shown in Table 1, indeed reveal that the correlation contributions to these molecular parameters do converge more rapidly for the "highly multireference" BeOBe example.

Results for a selected and somewhat chemically wider range of small molecules are documented in Table 2. The series of 10 electron hydrides ranging from the borohydride anion $BH_4^-$ to HF all contain exclusively single bonds and possess electronic wave functions that are dominated by a single Slater determinant. All of these systems present comparably simple challenges to the treatment of electron correlation. The differences between all five of the diagnostics listed in the table are too small to form the basis of any conclusions, but the magnitude of the values serves as an indicator of what values might be associated with "easy" molecules. Following these simple hydrides are the isoelectronic BN and $C_2$ species, both of which have very large $T_2$ amplitudes and whose singlet electronic ground states are well-known challenges in quantum chemistry. For these two, there is some disagreement as to which is the more difficult case; the $T_1$ diagnostic is larger for BN, and $C_2$ has the larger doubles amplitude. The proposed diagnostic favors the former conclusion, which is supported by the associated equilibrium geometries.[31]

Somewhat less difficult but still challenging are the isoelectronic series $N_2$, CO, and BF. Again, these molecules show relatively similar behavior vis-à-vis the $T_1$ diagnostic and largest $T_2$ amplitudes, with the relative behavior of the former quite similar to that of the new diagnostic. Finally, for ozone, which has long been known to present difficulties to quantum chemistry, the new diagnostic takes on values quite similar in magnitude to those of the 14-electron diatomics mentioned above, and the value of the diagnostic for ozone is greater than that of all other species studied here at the CCSDTQ level, except for BN, a finding that should not surprise any members of the computational chemistry community. In a forthcoming paper, the present work will be extended to methods containing non-iterative approximations to classes of excitation [i.e., CCSD(T),[32,33] CCSDT(Q),[34] etc.]; it will be interesting to see the variations of such results for the present series of molecules, with ozone a particularly salient example in this regard.

At the beginning of this research initiative, it appears that a formal shortcoming of CC theory, its non-Hermitian character, can be used to advantage in computational chemistry. Specifically, an easily computable manifestation of this characteristic is the asymmetry of the single-particle reduced density matrix in the molecular orbital representation. This work has shown that the extent of asymmetry correlates well with the rigor of calculations based on the associated wave function. As a result, users are provided with a diagnostic indicator of the propriety of a particular CC treatment. The diagnostic becomes larger when the problem becomes more difficult (similar to the usual behavior of the $T_1$ diagnostic and other measures) but *has the added property that it becomes smaller as the quality of the calculation is improved*. The present letter shows this correlation for a few simple cases studied with the CCSD, CCSDT, and CCSDTQ methods.

A reviewer wondered if there would be any statistical correlation between the DAD diagnostics and the deviation from the exact (i.e., full CI and FCI) correlation energy. We were able to obtain full CI/cc-pVDZ correlation energies (Table S2 of the Supporting Information) for the molecules in Table 2; for $O_3$, where this was unfeasible, we substituted a very close additivity approximation $CCSDTQ(5)_\Lambda$/cc-pVDZ + $CCSDTQ56(7)_\Lambda$/cc-pVDZ(no d) − $CCSDTQ(5)_\Lambda$/cc-pVDZ(no d). Correlation between the DAD index and the discrepancy from full CI across all of the molecules is quite weak. However, for individual molecules, going from CCSD to CCSDT to CCSDTQ, the DAD diagnostic exhibits a clear linear correlation with the residual error relative to full CI (see Figure 2; numerical values are given in Table S2 of the Supporting Information). The exception that proves the rule is the pathological BN diatomic, and even there, the trends run parallel.

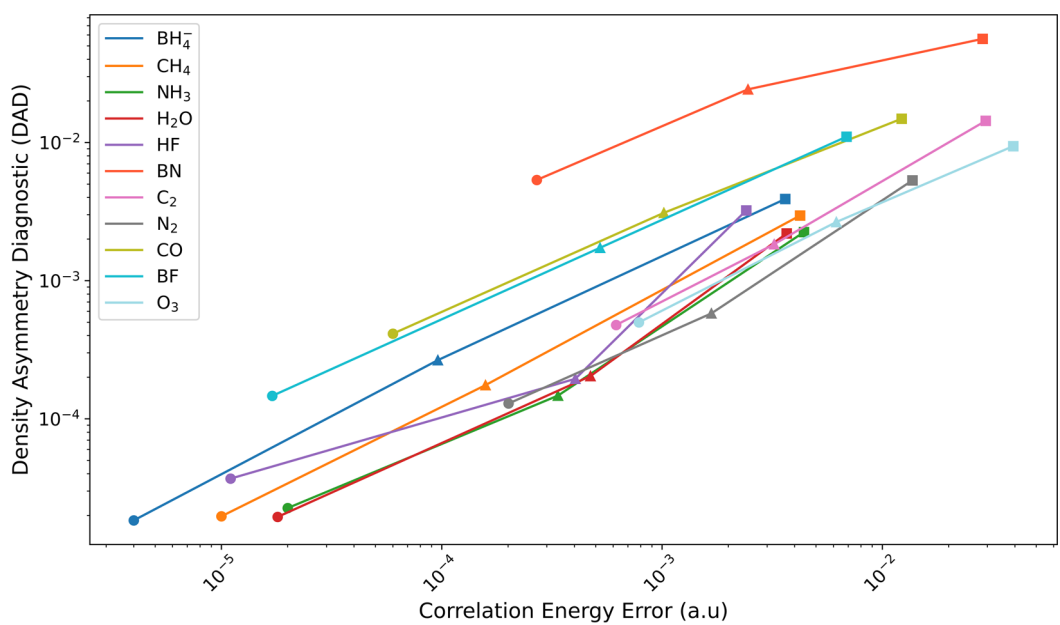

**Figure 2.** Density Asymmetry Diagnostic versus Correlation Energy Error for the Table II molecules in the frozen-core approximation at CCSD (squares), CCSDT (triangles) and CCSDTQ (circles) levels of theory using the cc-pVDZ basis set.

The same reviewer queried how DAD compares to other known static correlation diagnostics. A large number of such were evaluated in refs 17 and 35 for the closed-shell molecules in the W4-11 thermochemical benchmark;[36] aside from the classic $T_1$ diagnostic,[14] these included the matrix norm-based $D_1$ and $D_2$ diagnostics,[37,38] the percentage of parenthetical triples in the molecular total atomization energy %TAE[(T)],[40] the percentage of correlation energy in the same %TAE$_{corr}$ = 100% − %TAE[HF][40] as well as various measures based on the natural orbital occupation numbers, such as the $M$ diagnostic[39] and Matito's $I_{\overline{ND}}$ and $I_{\overline{ND}}^{max}$ diagnostics,[18] plus two DFT-based diagnostics introduced in ref 35: the percentage of exchange in the DFT atomization energy (%TAE[X]) and the difference between the exchange contributions from Hartree−Fock and self-consistent DFT orbitals.

It was already shown in ref 35 that principal component analysis (or indeed, simple visual "blocking" of the Pearson correlation matrix between the variables) reveals that all diagnostics cluster into three groups: (a) those based on single excitation amplitudes, (b) those based on double-excitation amplitudes or natural orbital occupations (including the correlation entropy[41]), and (c) pragmatic energy-based diagnostics, such as %TAE[(T)] or %TAE[X].

We added the DAD diagnostic to the data set as well as the two diagnostics proposed in ref 18. The spreadsheet is supplied in the Supporting Information, where in Table S3, we also present the coefficients of determination $R^2$ between pairs of diagnostics. Our expanded analysis unambiguously places DAD in cluster (A).

Note that molecules like O$_3$ and BN, for both of which $T_2^{max}$ (the largest double substitutions amplitude) is quite large, have very different DAD values, and this difference persists even for CCSDT and CCSDTQ. For the molecules in Table 2, Table S4 of the Supporting Information presents root mean square (RMS) and maximum values of single, double, triple, and quadruple substitution amplitudes. However, BN presents not only much larger $T_1^{max}$ but also $T_3^{max}$ and $T_4^{max}$ than O$_3$.

Further research planned for this area includes CC methods with non-iterative treatment of higher excitations [e.g., CCSD(T) and CCSDT(Q)], open-shell molecules and excited or other states accessed with the equation-of-motion (EOM) variants of the CC theory, and extensions that include the two-particle reduced density matrix (which is also not symmetric). We are confident that the diagnostic utility of the density asymmetry will carry over to these other cases and feel that this measure will be found useful and informative by those using CC calculations in their research.

## ■ ASSOCIATED CONTENT

**ⓈSupporting Information**

The Supporting Information is available free of charge at https://pubs.acs.org/doi/10.1021/acs.jpclett.5c00885.

Tables S1−S4 and Figure S1 (PDF)

Non-dynamical correlation diagnostics for molecules in closed-shell subset of W4-11 (XLSX)

## ■ AUTHOR INFORMATION

**Corresponding Author**

**Jan M. L. Martin** − *Quantum Theory Project, Department of Chemistry, University of Florida, Gainesville, Florida 32611, United States; Department of Molecular Chemistry and Materials Science, Weizmann Institute of Science, 7610001 Reḥovot, Israel;* ⓞ orcid.org/0000-0002-0005-5074; Phone: +972-8-9342533; Email: gershom@weizmann.ac.il; Fax: +972-8-9343029

**Authors**

**Kaila E. Weflen** − *Quantum Theory Project, Department of Chemistry, University of Florida, Gainesville, Florida 32611, United States;* ⓞ orcid.org/0009-0003-2483-8692

**Megan R. Bentley** − *Quantum Theory Project, Department of Chemistry, University of Florida, Gainesville, Florida 32611, United States;* orcid.org/0009-0007-6711-0051

**James H. Thorpe** − *Department of Chemistry, Southern Methodist University, Dallas, Texas 75275, United States;* orcid.org/0000-0002-9258-006X

**Peter R. Franke** − *Quantum Theory Project, Department of Chemistry, University of Florida, Gainesville, Florida 32611, United States;* orcid.org/0000-0001-9781-3179

**Devin A. Matthews** − *Department of Chemistry, Southern Methodist University, Dallas, Texas 75275, United States;* orcid.org/0000-0003-2795-5483

∥**John F. Stanton** − *Quantum Theory Project, Department of Chemistry, University of Florida, Gainesville, Florida 32611, United States;* orcid.org/0000-0003-2345-9781

Complete contact information is available at:
https://pubs.acs.org/10.1021/acs.jpclett.5c00885

**Notes**
The authors declare no competing financial interest.
∥Senior author: Deceased March 21, 2025.

## ■ ACKNOWLEDGMENTS

The surviving authors mourn the loss of their great mentor, colleague, and friend John F. Stanton, who passed away in the midst of the submission process of this manuscript. The authors express their deepest appreciation of John F. Stanton and what he was to each of them. He will be sorely missed. May his memory be blessed. The authors dedicate this work to the memory of Jiří Čížek, the pioneer who migrated the field of the coupled-cluster theory to quantum chemistry in 1966. That this method is still the focus of research testifies to his vision, and his recent parting has left the community without one of its great heroes. This work was initiated when Jan M. L. Martin was a sabbatical visitor at the Quantum Theory Project in the spring of 2024. It was supported by the U.S. Department of Energy, Office of Basic Energy Sciences, under Award DE-SC0018164 (to John F. Stanton), and the U.S. National Science Foundation under Grant CHE-2143725 (to Devin A. Matthews). Kaila E. Weflen acknowledges the support of the Chemical Physics Undergraduate Research Scholarship given by the Quantum Theory Project at the University of Florida. James H. Thorpe acknowledges support from the SMU Moody School of Graduate and Advanced Studies. Finally, we have benefited from useful discussions with P. B. Changala (JILA, Boulder), Gregory H. Jones (University of Florida), Peter R. Taylor (Tianjin U.) and Edit Mátyus and Attila Császár (ELTE, Budapest).